\documentclass[12pt]{article}
\usepackage{psfig,epsfig,graphics}
\usepackage{amssymb}

\begin{document}

\begin{center}
{\huge{An agent-based model to rural-urban migration analysis}}
\end{center}

\begin{center}
{Jaylson J. Silveira}\footnote{jaylson@fclar.unesp.br}\\
{Depto de Economia, Universidade Estadual Paulista - UNESP\\
 Araraquara, SP, Brazil}\\
\end{center}

\begin{center}
{Aquino L. Esp\'{\i}ndola}\footnote{aquino@if.uff.br}\\
{Instituto de F\'{\i}sica, Universidade Federal Fluminense\\
Niter\'oi, RJ, Brazil}\\
\end{center}

\begin{center}
{T. J. P. Penna} \footnote{tjpp@if.uff.br} \\
{Instituto de F\'{\i}sica, Universidade Federal Fluminense\\
Niter\'oi, RJ, Brazil}\\
\end{center}

\vspace{0.6cm}

\begin{abstract}
In this paper we analyze the rural-urban migration phenomena as it is usually observed in economies which are in the early stages of industrialization. The analysis is conducted by means of a statistical mechanics approach which builds a computational agent-based model. Agents are placed on a lattice and the connections among them are described via an Ising like model. Simulations on this computational model show some emergent properties that are common in developing economies, such as a transitional dynamics characterized by continuous growth of urban population, followed by the equalization of expected wages between rural and urban sectors (Harris-Todaro equilibrium condition), urban concentration and increasing of \textit{per capita} income.
\end{abstract}


\section{Introduction}

\par Economic development generates significant structural transformations, such as changes in the demographic condition and the production structure. As mentioned in chapter 10 of ref. \cite{ray} the most important structural feature of developing economies is the distinction between rural and urban sectors. The agriculture plays a key role in the development of the urban sector of an economy: it provides both the food surplus that enables the urban sector to survive, and the supply of labor to the expanding industrial urban sector. As suggested in chapter 5 of ref. \cite{willianson}, the fundamental part of the transformation from mostly dispersed rural and agrarian countries in a more concentrated urban and industrial is the flux of a large number of individuals through migration from rural areas to urban areas.

\par In this paper we examine the rural-urban migration phenomena which takes place during the industrialization process. The analysis is carried out by using an agent-based computational model which aims to describe some of the main structural features of a developing economy. We look at the rural-urban migration as a discrete choice problem, which allows us to formalize the migration process by using an Ising like model. We modelled the migratory decision in the usual manner considering the force exerted by the difference of earnings between the sectors. Moreover, it is also included a new factor, the influence from neighbors like in the Ising model.

\par The paper is organized as follows. Section \ref{economicsetting} describes the economic setting, i.e., the typical dual economic structure (rural \textit{versus} urban sector) in industrializing countries. In section \ref{migratoryprocess} the migration process is modelled within a statistical mechanics framework. Section \ref{simulation} presents the simulations and the main results. Finally, section \ref{conclusion} shows the concluding remarks.

\section{The Economic Setting}
\label{economicsetting}

\par Let us consider an economic system formed by two sectors, one rural and the other urban. The main differences between these sectors are the sort of goods they produce, the technologies used by firms and the framework of wage determination. Such a dual structure is typically used by the economic literature which investigates the rural-urban migration phenomena \cite{ray}\cite{harristodaro}\cite{todaro}\cite{bardhan}\cite{ranis}. The basic features of the dual economy will be described in subsection \ref{urbansector} and \ref{ruralsector}. Subsection \ref{macrostate} shows how the equilibrium macrostate of the economic system is determined.

\subsection{The urban sector}
\label{urbansector}

\par The urban productive sector is formed by firms specialized in the production of manufacturated goods. The output of the $ith$ firm $Y_{mi}$ depends positively on both the amount of employed workers $N_{mi}$ and the effort $\varepsilon$, spent by each worker to perform his job. Based on the classical rural-urban migration theory \cite{harristodaro}\cite{todaro}, we assume that the stock of capital during the analysis period is given.  Supposing a standard geometrical functional form \cite{day}, the production function of the manufacturing firm is the well-known Cobb-Douglas\footnote{The modelling of employment and wage determination of the urban sector is based on the efficiency-wage approach. See chapter 10 of ref. \cite{romer} and section V of ref. \cite{mankiwromer} for further details.}  
 
\begin{equation}
Y_{mi}=A_m\left(\varepsilon N_{mi}\right)^\alpha,
\label{ymi}
\end{equation}
where $0<\alpha<1$ and $A_m>0$ are parametric constants.

\par By using the functional form originally suggested by Summers \cite{summers} and slightly modified by Romer \cite{romer}, the urban worker's effort can be defined as a function of the real wage paid by the manufacturing firm, the urban unemployment rate $u$ and the alternative wage $w_m$, which is paid by other firms of the same sector. Then the effort function is given by

\begin{equation}
\varepsilon=
\left\{\begin{array}{l}
\left[\frac{w_{mi}-(1-bu)w_m}{(1-bu)w_m}\right]^\eta,\quad\mbox{if}\quad w_{mi}>(1-u)w_m, \\
0,\quad\mbox{otherwise,}\\
\end{array}
\right.
\label{effort}
\end{equation}
where $0<\eta<1$ and $b>0$ are parametric constants.
 
\par The $Z_m$ manufacturing firms which form the manufacturing sector seek to maximize their real profits, measured in units of the manufacturated good, by choosing wages and employment freely. Given eq. (\ref{ymi}) and eq. (\ref{effort}) the real profit of $i$th manufacturing firm is

\begin{equation}
 A_m\left[\left(\frac{w_{mi}-(1-bu)w_m}{(1-bu)w_m}\right)^\eta N_{mi}\right]^\alpha - w_{mi}N_{mi}.
\label{maxprofit}
\end{equation}

\par The maximization condition of eq. (\ref{maxprofit}) can be found using the first-order condition for a maximum, which result is\footnote{The second-order condition for a maximum is also satisfied.}

\begin{equation}
w_{mi}=\frac{(1-bu)w_m}{1-\eta}
\label{wmi}
\end{equation}
and
\begin{equation}
N_{mi}=\left[\frac{\alpha A_m\eta^{\alpha\eta}(1-\eta)^{1-\alpha\eta}}{(1-bu)w_m}\right]^{\frac{1}{1-\alpha}}.
\label{nmimax}
\end{equation}

\par In equilibrium, all these firms choose the same wage \cite{summers}\cite{romer}, i.e., $w_{mi}=w_m$ $(i=1,2,3,...,Z_m)$. Then, from equation (\ref{wmi}) the equilibrium urban unemployment rate is
 
\begin{equation}
u=\frac{\eta}{b}.
\label{u}
\end{equation}

By definition, the urban unemployment rate is the ratio between the number of unemployed workers and the urban population $(N_u-Z_mN_{mi})/N_u,$ where $N_u$ is the amount of workers localized in the urban sector. The previous definition must be consistent in each period to the equilibrium value of (\ref{u}). The employment level of the manufacturing firm which obeys this consistency condition is obtained equaling the equilibrium in eq. (\ref{u}) to the previous definition:

\begin{equation}
N_{mi}=\left(1-\frac{\eta}{b}\right)\frac{N_u}{Z_m}.
\label{nmi}
\end{equation}

\par Taking eq. (\ref{effort}), evaluated in the equilibrium, and eq. (\ref{nmi}) and replace both in eq. (\ref{ymi}), the aggregated production of the manufacturing sector, $Z_mY_{mi}$, is given by
 
\begin{equation}
Y_m=\xi_1{N_u}^\alpha,
\label{ym}
\end{equation}
where $\xi_1=A_mZ_m^{1-\alpha}\left[\left(\frac{\eta}{1-\eta}\right)^\eta\left(1-\frac{\eta}{b}\right)\right]^\alpha$.

\par By using eqs. (\ref{nmimax}), (\ref{u}) and eq. (\ref{nmi}) one can obtain the equilibrium wage of the manufacturing sector:

\begin{equation}
w_m=\xi_2{N_u}^{\alpha-1},
\label{wm} 
\end{equation}
where $\xi_2=\alpha A_m\left(\frac{\eta}{1-\eta}\right)^{\alpha\eta}\left[\left(1-\frac{\eta}{b}\right)\frac{1}{Z_m}\right]^{\alpha-1}$.

\par Given the parametric constants that specify the technology and the effort sensitivity to wage, as well as the size of the manufacturing productive sector, it is possible to see that the equilibrium of the urban sector depends directly upon the urban population $N_u$.

\subsection{The rural sector}
\label{ruralsector}

\par In the rural sector the farm $i$ produces an agricultural output $Y_{ai}$ by employing an amount of workers $N_{ai}$. The output is obtained by using a Cobb-Douglas production function \cite{day}

\begin{equation}
Y_{ai}=A_a\left(N_{ai}\right)^\phi ,
\label{yai}
\end{equation}
where $0<\phi<1$ and $A_a>0$ are parametric constants. We suppose that both the land endowment and the stock of capital of the farm are given during the period of analysis as assumed by refs. \cite{harristodaro} and \cite{day}.

\par Differently from the urban sector, farms are price-takers and the real wage is adjusted up to the point in which there is no unemployment in this sector \cite{harristodaro}\cite{todaro}. This implies that the rural population will match the aggregated employment in the rural sector. Therefore, the equilibrium employment level of the farm $i$ is

\begin{equation}
N_{ai}=\frac{N-N_{u}}{Z_a},
\label{nai}
\end{equation}
where $Z_a$ is the amount of farms which constitute the agricultural sector and $N$ is the total population of the economic system.

\par From eq. (\ref{yai}) and eq. (\ref{nai}) the aggregated production of the rural sector, $Z_aY_{ai}$, is

\begin{equation}
Y_a=\xi_3\left(N-N_u\right)^\phi ,
\label{ya}
\end{equation}
where $\xi_3=A_aZ_a^{1-\phi}$.

\par Thus, the profit maximizing of the farms imply that the rural real wage expressed in units of the manufactured good becomes equal to the marginal product of agricultural labor in units of manufacturing good\footnote{This marginal product is the derivate of the production function, eq. (\ref{yai}), with respect to $N_{ai}$ multiplied by $p$.}:

\begin{equation}
w_a=\xi_4p\left(N-N_u\right)^\phi,
\label{wa}
\end{equation}
where $\xi_4= \left( A_a\phi / Z_a^{\phi-1}\right)$ and $p$ is the price of the agricultural good expressed in units of the manufactured good. 

\par Like in the urban sector, the equilibrium state of the rural sector depends on the urban population, as the size of total population of the economy is fixed.

\subsection{The macrostate of economic system}
\label{macrostate}

\par As proposed by Harris and Todaro \cite{harristodaro}, the terms of trade between the rural and urban sectors, measured by the price $p$, depend on the relative scarcity of agricultural and manufacturated goods. This can be measured by the ratio $Y_m/Y_a$. The greater this ratio the greater will be the scarcity of agricultural good, which implies an increase of the agricultural good price in units of manufacturated good. Formally, given the urban population, the equilibrium relative price of the agricultural good is\footnote{In the literature on rural-urban migration it is usual, because of analytical simplicity, to consider $p$ constant \cite{day}. This is true in the special case when $\gamma=0$ in eq. (\ref{p}).}

\begin{equation}
p=\rho\left(\frac{Y_m}{Y_ a}\right)^\gamma 
\label{p}
\end{equation}
where $\rho>0$ and $\gamma>0$ are parametric constants.

\par Therefore, given the size of urban population, by using equations (\ref{u}-\ref{wm}) one can calculate the state of urban sector. Likewise, the rural sector state is determined by means of equations (\ref{nai}-\ref{p}). The equilibrium state of both sectors will be modified if a migratory flux changes the population distribution of the economic system.

\section{Migratory process: a statistical mechanics approach}
\label{migratoryprocess}

\par As argued by Harris and Todaro \cite{harristodaro}\cite{todaro}, individuals take their decisions of migrating or not by considering the differential of expected wages between their present sector and the sector they intend to go. However, other authors have taken into account additional reasons.
Based on the formalization from statistical mechanics applied to socioeconomic phenomena \cite{brock}\cite{durlauf}, in this section we propose an agent-based computational model to describe the rural-urban migratory process. This model is focused on the influence that individuals suffer in the reference group that they are included. The emergent properties will be analyzed taking into account the standard effect of labor allocation Harris-Todaro mechanism, which is based on the expected differential wages between sectors. This analysis will also be concerned on the effect of social neighborhood, often mentioned by other authors but not yet formalized.

\par The main feature of the decision process is that each worker reviews his sectorial location after a period of time spent in that sector. We exclude, by assumption, the possibility that the worker may simultaneously supply his labor force to both sectors. Thus, only two choices are admitted: stay in the sector in which he was during previous periods or migrate.

\par In order to model the migration process by allowing only discrete choices, each worker has its state defined by $\sigma_i\in\{-1,+1\}$, where $\sigma_i=-1$ means that the worker is at the rural sector; otherwise, $\sigma_i=+1$, representing the urban sector. 

\par In our model, during the decision process, explicit and observable incentives are taken into account by each potential migrant. This is called a deterministic private utility \cite{brock}\cite{durlauf}, given by

\begin{equation}
U_i=H(t)\sigma_i,
\label{ui}
\end{equation}
where $H(t)=k\omega_e$, $k>0$ is a parametric constant and $\omega_e$ is the expected urban-rural differential wage. The expected urban-rural wages in function of $H(t)$ are specified as follows.

\par Jobs are allocated at random when manufacturing firms are faced with more applicants than jobs avaliable \cite{harristodaro}\cite{bardhan}. It means that in each time step all urban workers have the same probability to find an urban job. Under such a hypothesis, the term $(1-u)$ is the probability of an urban worker to obtain a job. Hence, $(1-u)w_m$ is the expected urban wage. Assuming that the rural wage is perfectly flexible there is no unemployment in the rural sector. Then, the probability to find a job in the rural sector is $1$. Therefore, the rural wage $w_a$ is the same as the expected wage in this sector. In sum, the expected differential of wage between urban and rural sectors is

\begin{equation}
\omega_e=(1-u)w_m-w_a.
\label{omegae}
\end{equation}

\par Besides, the worker $i$ is also under the influence of other workers,  his social neighborhood \cite{gustavo}, denoted by $n_i$. The measure of such influence, that is, the deterministic social utility \cite{brock}\cite{durlauf}, is given by

\begin{equation}
S_i=J\sum_{j\in n_i}\sigma_i\sigma_j,
\label{spin}
\end{equation}
where $J>0$ is a parametric constant. The term $J$ represents the interaction weight which relates the worker $i$'s choice to the neighbor $j$'s choice. This is assumed to be nonnegative, by representing the hypothesis that the worker seeks to conform to the behavior of his neighbors \cite{durlauf}. The interactions among neighbors are assessed in the workers' nearest neighbors or in the next nearest neighbors.

\par Then, following references \cite{brock} and \cite{durlauf}, we assume that payoff of worker $i$, which is his deterministic total (private and social) utility can be obtained replacing eq. (\ref{omegae}) in eq. (\ref{ui}) and summing with eq. (\ref{spin}):

\begin{equation}
{\mathcal{H}}_i=k\left[(1-u)w_m-w_a\right]\sigma_i+J\sum_{j\in n_i}\sigma_i\sigma_j.
\label{hi}
\end{equation}

\par Therefore, this system can be described by the well-known ferromagnetic Ising model, in the presence of an external time-dependent magnetic field:

\begin{equation}
{\mathcal{H}}=-H(t)\sum_{i=1}^N\sigma_i - J\sum_{<ij>}\sigma_i\sigma_j.
\label{ising}
\end{equation}

\par In each time step, each worker reviews his decision about the sectorial location with probability $a$, called activity \cite{thadeu}. Then, there is a part of the population that reviews their decisions and becomes potential migrants.

\par The potential migrant $i$ becomes an actual migrant depending on the comparison between his deterministic total utility  ${\mathcal{H}}_i$ and his non observable and idiosyncratic motivations $\mu_i$, called random private utility \cite{durlauf}. The term $\mu_i$ represents the net difference between the random private utilities that the potential migrant assigns to the sector he intends to move and his present sector. 

\par In each period, if $\mu_i > {\mathcal{H}}_i$, the potential migrant $i$ becomes an actual migrant; otherwise, this does not happen. Supposing that $\mu_i$ is a random variable logistically distributed \cite{brock}\cite{durlauf}, the probability that the potential migrant effectively migrates is given by a cumulative distribution:

\begin{equation}
Pr_i=\frac{1}{1+e^{-\beta {\mathcal{H}}_i}},
\label{prob}
\end{equation}
where $\beta > 0$ is a parametric constant that in this context measures the heterogeneity of workers concerning to the migration propensity. Equation (\ref{prob}) is a measure of the probability that a worker $i$, who is reviewing his location strategy, stays in the sector that he is localized at that time. The higher his deterministic total utility, eq. (\ref{hi}), the higher the probability that no change will take place.

\section{Simulation}
\label{simulation}

\par To carry out the simulation of the economic system described in the previous sections, each worker is placed in one of the sites of a square lattice. The state of each site (worker) is set as mentioned before: $\sigma_i=+1$ for urban workers and $\sigma_i=-1$ for rural ones. It is important to emphasize that the state of these sites represent the sectorial allocation of each worker, i.e., whether an individual is suppling his labor force in the urban or rural sector. It means that the coordinates of the lattice sites are not related to spatial distribution of workers.

\par To set up the initial state of the system, all workers are randomly distributed in the lattice. At time $t=0$ there is the initial condition that $20\%$ of the population is urban. In other words, initially, $20\%$ of the sites will be assigned with $\sigma_i=+1$ and the remaining $80\%$, $\sigma_i=-1$. The reason for this initial distribution is because these are the values which have usually been observed in developing countries before the urbanization process initiates.

\par The next step in the simulation is to calculate the equilibrium state variables of the urban sector, by using eqs. (\ref{u}-\ref{wm}), and of the rural sector by using eqs. (\ref{nai}-\ref{p}). Since the state variables of both sectors are known, it is necessary to define the amount of workers that will review their sectorial location, i.e. those one who will become potential migrants. To do this, it is assumed that the probability that a worker will become a potential migrant is given by the activity $a$, as defined by Stauffer and Penna \cite{thadeu}. All those selected as potential migrants will have their private utility calculated by eq. (\ref{hi}).

\begin{figure}[hbt]
\centerline{\psfig{file=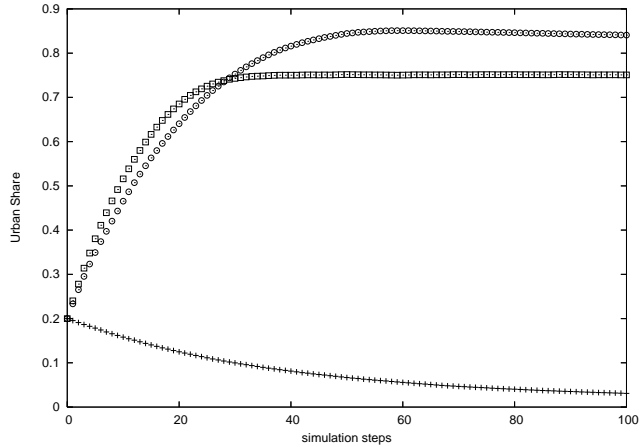,height=6.0cm,angle=0}}
\caption{Proportion of workers at urban sector as function of time for three different set of parameters $J$ and $k$. Circles: $(J>0$, $k>0)$; Squares: $(J=0$, $k>0)$; Crosses: $(J>0$, $k=0)$.}
\label{nut}
\end{figure}

\par In order to conclude the reviewing process, the probability defined in eq. (\ref{prob}) is assessed. Then, a random number $rn\in [0,1]$ is generated from an uniform distribution. If $rn>Pr$, then the potential migrant becomes an actual migrant; otherwise, no change takes place. 

\par As soon as the potential migrants end their reviewing process, a new sectorial distribution is obtained. Knowledge of the new urban population allows the macrostate of the economic system to be reset. Therefore, the state variables of both sectors have to be calculated again. The whole procedure described above will be repeated as many times as we set in the simulation. The stopping criteria used by us is halting the simulation some steps after the moment when the system reaches equilibrium.

\par Figure \ref{nut} shows the proportion of workers in the urban sector $n_u\equiv \frac{N_u}{N}$, from now on called urban share, plotted in three different combination of the parameters $J$ and $k$. It is necessary to remind that the parameters $J$ and $k$ adjust the instensity of the deterministic private utility, eq. (\ref{ui}), and deterministic social utility, eq. (\ref{spin}), respectively. From top to bottom the set of parameters used in the plotting are $(J>0$, $k>0)$, $(J=0$, $k>0)$ and $(J>0$, $k=0)$.

\par Firstly, consider the case $(J=0$, $k>0)$ plotted in  Fig. \ref{nut}. In this case, the review conducted by the agents is guided only by the deterministic private utility, which in turn depends on the expected urban-rural difference of wages. As in models of classical theory of migration \cite{harristodaro}\cite{todaro}, when the expected urban wage is higher than the rural wage, it implies in a continuous growth of the urban share, as well as a relatively fast convergence towards the equilibrium.

\begin{figure}[hbt]
\centerline{\psfig{file=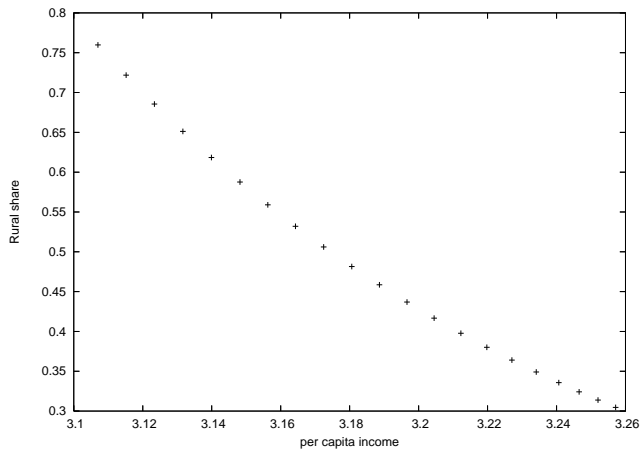,height=6.0cm,angle=0}}
\caption{Rural share as function of \textit{per capita} income in units of manufacturated good.}
\label{income}
\end{figure}

\par Secondly, consider the case where both effects are taken into account, $(J>0$, $k>0)$. Like the previous case, the rural-urban migratory process occurs again, however, the system reaches a higher value of the equilibrium urban share, though it takes more time for such outcome. This difference is caused by the parameter $J>0$, what means that the influence of the social neighborhood is considered. To better understand this behavior, it should be reminded that the process of sectorial position revision depends on the deterministic private utility and the social private utility. Then, when $J>0$ the influence of social neighborhood is being exerted, i.e., each worker attempts to adjust his choice according to the sectorial position of his neighbors. The existence of such an influence causes two different effects during the process of convergence towards equilibrium. In the first moment, when the neighborhood are mainly rural, the influence from neighbors slows the rural-urban migratory flux, increasing the time necessary to reach equilibrium. In the second moment, when the neighborhood become mainly urban, the influence reinforces the attraction from the high expected urban wage, leading to higher equilibrium urban share. 

\par Finally, we consider the case $(J>0$, $k=0)$, with only neighborhood effects shown. In this case, the potential migrants consider only the sectorial position of the neighborhood and do not  take into account the expected differential of wages. The pure effect due from neighborhood leads to the extinction of the urban sector. This is not an empirically important case, as it has not been observed in developing economies.

\par In Figure \ref{income}, another important feature caused by the migratory dynamics is the expansion of \textit{per capita} income $(Y_m+pY_a)/N$. This result matches to the economic data in which in countries with high \textit{per capita} income the proportion of the population living in rural area is low \cite{ray}.

\par
\begin{figure}[hbt]
\centerline
{
$\stackrel{\psfig{file=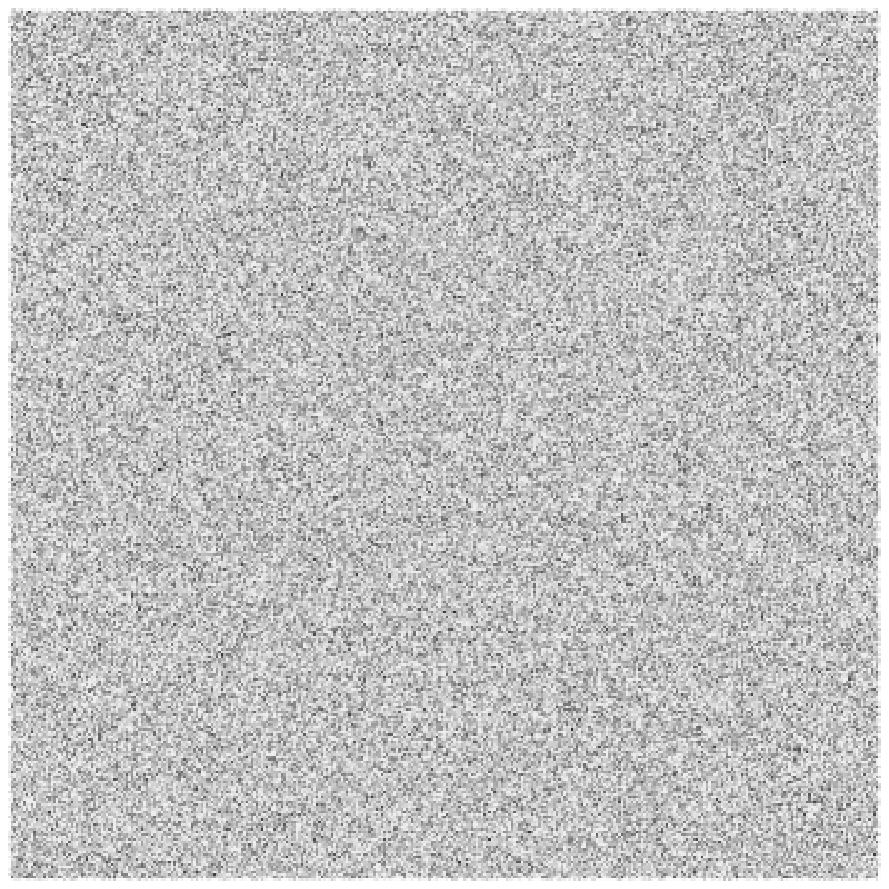,height=6.0cm,angle=0}} {(a)} $
$\stackrel{\psfig{file=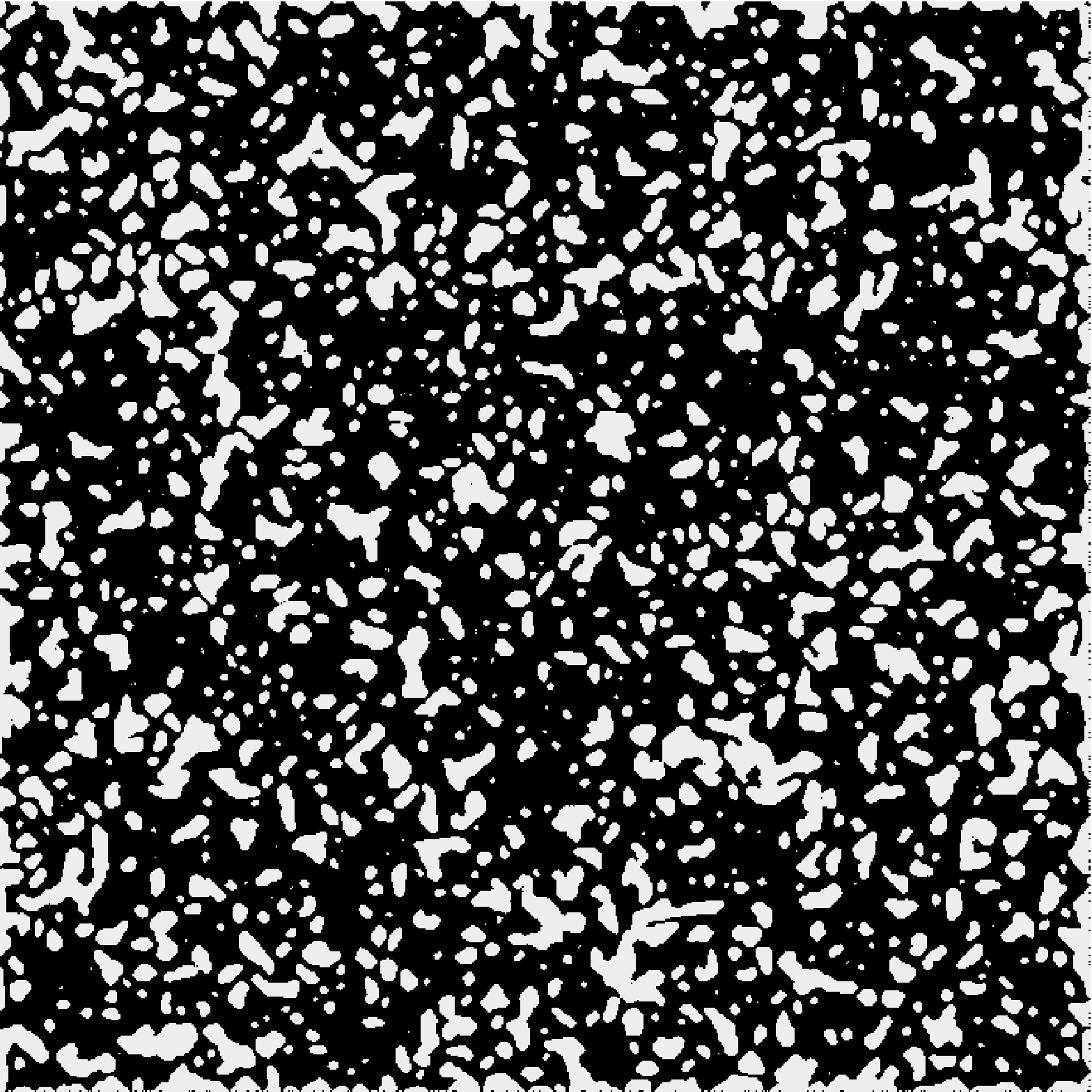,height=6.0cm,angle=0}} {(b)} $
}
\caption{Distribution of workers in the lattice. Black clusters represent urban workers ($\sigma_i=+1$) and white clusters are rural workers ($\sigma_i = -1$). Fig. (a) is the random distribution in the initial state of the system. Fig. (b) is the equilibrium distribution where clusters due the sectorial neighborhood can be seen.}
\label{lattice}
\end{figure}

 In the initial state of the system the configuration was randomly set with 20\% of the sites assigned $\sigma_i=+1$, urban workers, and the rest $\sigma_i=-1$, rural workers as shown in Fig. \ref{lattice}a. The final state of the dynamics by using $(J>0$, $k>0)$ can be visualized in Figure \ref{lattice}b. Now the infinite cluster is formed by sites $\sigma_i=+1$ representing the urban concentration caused by the migratory process. Several others clusters are formed by sites $\sigma_i=-1$.

\par
\begin{figure}[hbt]
\centerline{\psfig{file=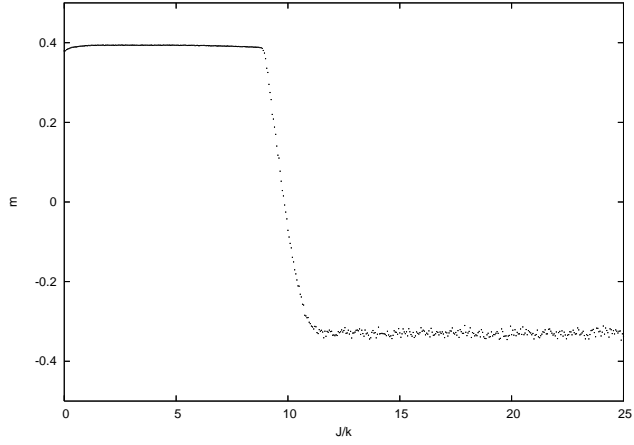,height=6.0cm,angle=0}}
\caption{Average magnetization $m$ as function of ratio $J/k$.}
\label{Jk1}
\end{figure}

\begin{figure}[hbt]
\centerline{\psfig{file=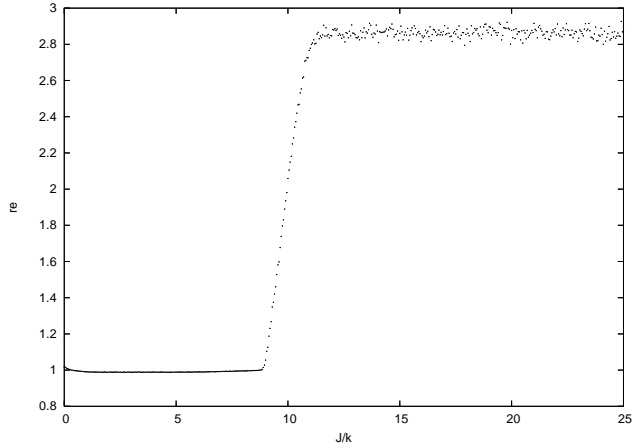,height=6.0cm,angle=0}}
\caption{Expected wages ratio $r_e$ as function of ratio $J/k$.}
\label{Jk2}
\end{figure}

 Figures \ref{Jk1} and \ref{Jk2} show the average magnetization $m=\sum\sigma_i/N$ and the expected wages ratio $r_e\equiv (1-u)w_m/w_a$, respectively. Both figures are plotted as function of the ratio $J/k$ ($k$ kept constant) measuring the relative intensity between these parameters.

\par Figure \ref{Jk1} has plotted in its vertical axis the average magnetization calculated during a period after the system have reached equilibrium. To values of $J/k$ less than the critical threshold the net magnetization is $m\cong 0.4$ representing an urban share about $n_u=0.70$. By increasing the ratio $J/k$ after this critical threshold the system goes to a new regime, changing completely its net magnetization.

\par Figure \ref{Jk2} is a plotting of expected wage ratio as function of $J/k$. To values $J/k\lesssim 9.0$ the ration is $r_e \cong 1.0$, what indicates that the expected urban wage and the rural wage converge to the same value. This property is known as Harris-Todaro equilibrium condition \cite{harristodaro}\cite{ray}\cite{bardhan}. Hence, in  a economic system where internal migration occurs freely the absolute difference between the rural and urban wages can persist if workers consider the possibility of unemployment. After the threshold $J/k>9.0$, $r_e$ has its maximum value around $2.8$ which shows that the urban expected wage is $2.8$ times greater than the rural wage. Even having this ratio increasing the value of the worker private utility, eq. (\ref{ui}), the equilibrium of the system is $m \cong -0.29$, i.e., a rural concentration of $64.9\%$. The explanation of this outcome is that after a given threshold the values of $J$ are in such a range that the social utility, eq. (\ref{spin}), is many times higher than the private utility. In other words, in such range, it does not matter if the expected wage is attractive in the urban sector because the strongest factor in the migration decision is the influence of the neighborhood, i.e., agents tend to mimic the behavior of other agents.

\par
\begin{figure}[hbt]
\centerline{\psfig{file=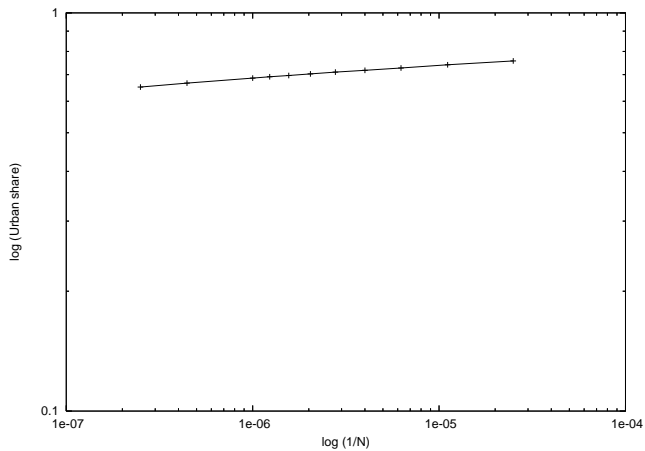,height=6.0cm,angle=0}}
\caption{Log-log plot of equilibrium urban share $n_u$ as function of the inverse of population size $1/N$.}
\label{nuN}
\end{figure}

 Simulations plotted in Figure \ref{nuN} indicate that when the size $N$ of the lattice increases the equilibrium urban share $n_u$ will change. For a given heterogeneity of the agents $\beta$, there is a power law relating equilibrium urban share and the inverse of lattice size. This can be formalized in the expression below

\begin{equation}
n_u = A\left(\frac{1}{N}\right)^\theta,
\label{linear}
\end{equation}
where $A$ and $\theta$ are constants which have to be estimated. To carry out the estimation of these constants we evaluated a linear regression of the log-linear version of eq. (\ref{linear}). In Table 1 one can find the estimation of the constants $A$ and $\theta$ based on data generated for five different values of $\beta$.

\par The estimation of the constants are approximately the same when using slightly different values of agent heterogeneity $\beta$. For example, by using any pair of constants $A$ and $\theta$ from Table 1, the estimation of equilibrium urban share by eq. (\ref{linear}) is $n_u=0.61$ for an economy with 50 million of workers.

\begin{center}
\begin{tabular}{ll|cc|c}
\hline
$\beta$ &       &$A$    &       &$\theta$\\
\hline
1.5     &       &1.064  &       &0.032\\
2.0     &       &1.061  &       &0.031\\
2.5     &       &1.061  &       &0.031\\
3.0     &       &1.064  &       &0.032\\
3.5     &       &1.066  &       &0.032\\
\hline
\end{tabular}
\end{center}
\begin{center}
{\bf{Table 1.}} Estimates of parameters $A$ and $\theta$ for different values of $\beta$.
\end{center}

\par
\begin{figure}[hbt]
\centerline{\psfig{file=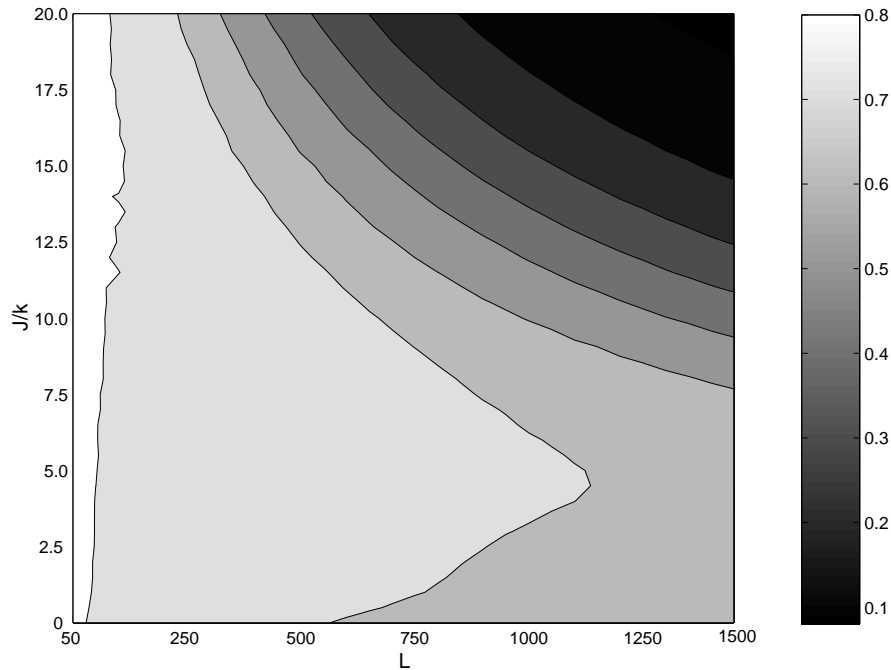,height=9.0cm,angle=0}}
\caption{Urban share $n_u$ as function of the ratio $J/k$ and the square lattice size $L$. Lighter areas correspond to higher $n_u$ and darker areas to lower $n_u$.}
\label{JN}
\end{figure}

 The effects of the ratio $J/k$, together with size of population $N=L^2$, are shown in Fig. \ref{JN}. The different values of equilibrium urban share are plotted in a grey scale. The first property observed in this figure is the existence of several phase states which depend on the values of $J/k$ and $N$. Each phase state is characterized by a constant equilibrium urban share. The topology of Figure \ref{JN} is in agreement with the results shown in Figs. \ref{Jk1} and \ref{nuN}, demonstrating that the properties of equilibrium macrostate depends on the combination of these parameters.

\begin{figure}[hbt]
\centerline{\psfig{file=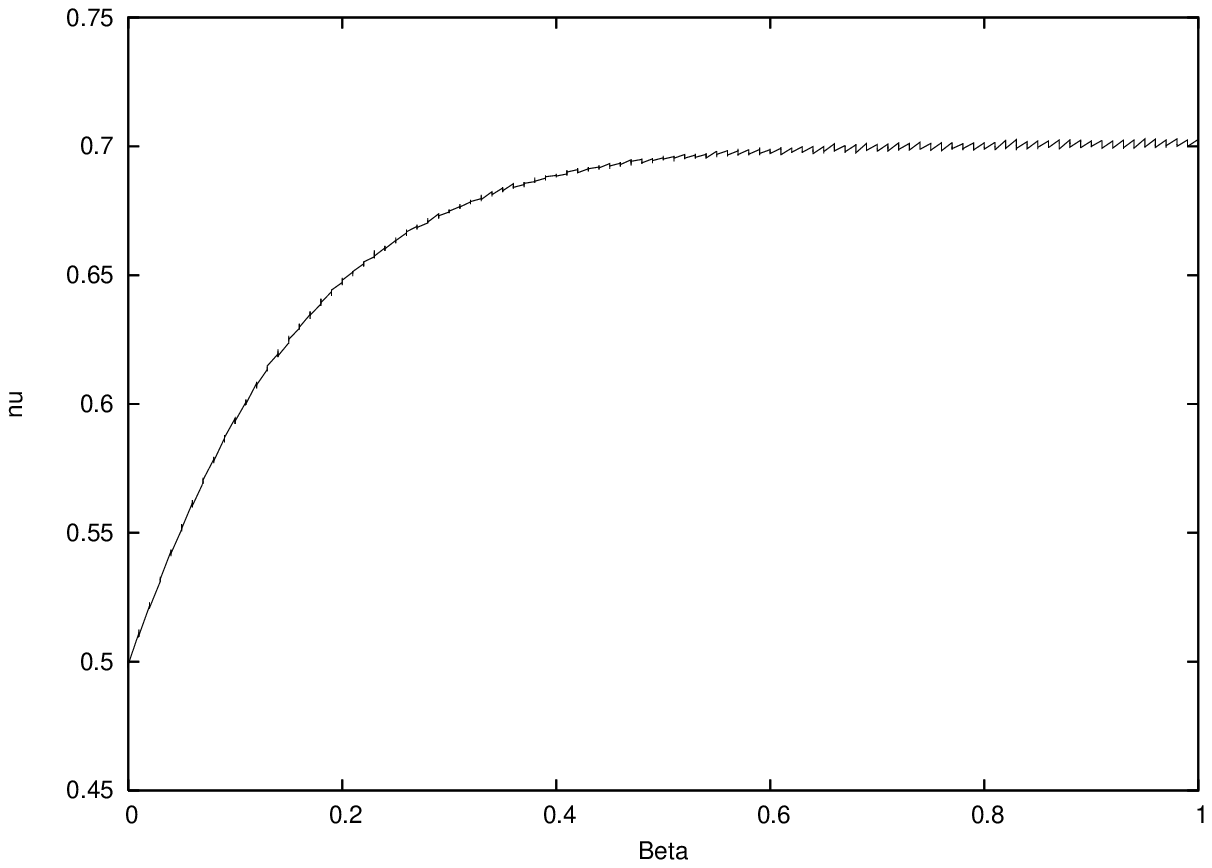,height=6.0cm,angle=0}}
\caption{Equilibrium urban share $n_u$ as function of the parameter $\beta$.}
\label{beta}
\end{figure}

\par  In Figure \ref{beta} is plotted the equilibrium urban share as function of the parameter $\beta$. For values of $\beta$ tending to zero the equilibrium urban share tends to 0.5 (or $m=0$), which implies in a null urban concentration (null average magnetization), even though there is an expected urban wage higher than the rural wage. In fact, eq. (\ref{prob}) shows that the smaller $\beta$ the higher the idiosyncratic and non-observed proportion of the worker's behavior related to the migration propensity. If $\beta = 0$, the choices  $\sigma_i = +1$ and $\sigma_i = -1$ have the same probability to occur being independent of the expected differential of wages. In sum, when the heterogeneity of the workers related to the decision of migration increases, the urban concentration will decline in the long run. On the other hand, when the heterogeneity of the agents decreases, i.e., $\beta$ increases, the equilibrium urban share is invariable after a threshold.

\begin{figure}[hbt]
\centerline{\psfig{file=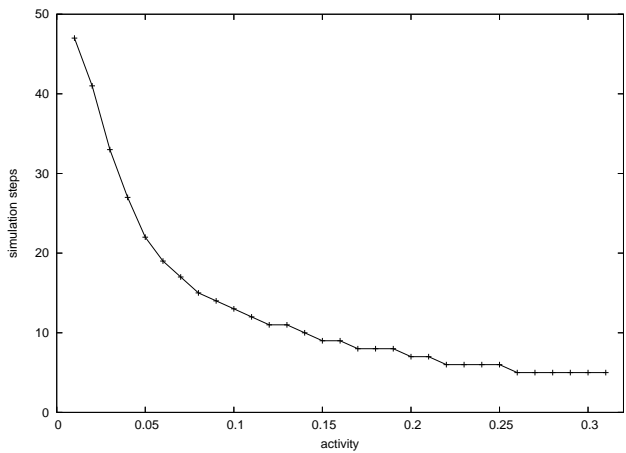,height=6.0cm,angle=0}}
\caption{Time to reach equilibrium as function of activity parameter $a$.}
\label{activity}
\end{figure}

\par The decision of migration is not taken simultaneously by all individuals. In order to simulate this behavior, the parameter called activity $a$ is used. It gives the probability that a worker will review the decision about his sectorial location. More specifically, $a$ represents the fraction of the population which will go through the reviewing process. This fraction of individuals is randomly selected and changes in each time step. In Fig. \ref{activity} variation of $a$ in different simulations shows that the time needed for the system to reach equilibrium is proportionally inverse to the value of the activity. Therefore, the time needed to reach the equilibrium state is strongly related to the amount of individuals which review their sectorial decision.

\newpage

\section{Conclusion}
\label{conclusion}

\par This paper has developed an agent-based computational model to analyse the rural-urban migration phenomena. The basic scenario was made of an economic system formed by two sectors, rural and urban, which differed in term of the goods produced, the production technology and the mechanism of wage determination.

\par By assuming the sectorial migration decision as discrete choice in a milieu of decentralized and non-coordinated decision-making, the rural-urban migration process was formalized as an Ising like model. The simulations showed aggregate regularities which indicates that decentralized migration decisions can lead to the emergence of equilibrium macrostates with features observed in developing economies. First, the simulation having an initial macrostate with population predominantly rural and expected urban wage higher than rural wage provoked a transitional rural-urban migratory dynamics, with continuous growth of the urban share. This is a key feature of the phenomena called in ref. \cite{willianson} as \textit{urban transition}.

\par Second, simulations also showed that, during the rural-urban migration process, the reduction of the rural share takes place together with the increasing of \textit{per capita} income of the economy. Such an inverse relation between rural share and \textit{per capita} income is one of the most robust facts detected in economic statistics \cite{ray}.

\par Third, the transitional rural-urban migratory dynamics converged towards an equilibrium macrostate. The features of this transitional dynamics and equilibrium are sensitive to the relative weight between private and social effects (utilities) as well as the degree of heterogeneity of agents concerning the migration propensity. When the social interaction component is relatively stronger and below a critical threshold the transitional dynamics towards equilibrium is delayed and reaches a higher equilibrium urban share. With a high heterogeneity of agents, $\beta\rightarrow 0$, this generates the end of the pulling force due the high expected urban wage what makes the system to reach an equilibrium macrostate with an urban share $n_u=0.5$. On the other hand, with a moderate heterogeneity of agents, $\beta>1$, the equilibrium urban shares will be set in a empirically reasonable range ($n_u\geq 0.6$).

\par The analysis shown in this paper suggests that a deeper investigation can still be carried out, which adopt alternative hypothesis mainly regarding the private and social utilities as well as other assumptions employed in our model.

\section*{Acknowledgments}

\hskip 0.5cm We would like to thank  Dietrich Stauffer, D. N. Dias, T. Lobo for their contributions and Dr. Renato P. Colistete for his comments. Jaylson J. Silveira acknowledges research grants from CNPq. Aquino L. Esp\'{\i}ndola thanks CAPES for the financial support. T. J. P. Penna thanks CNPq for the fellowship.


\begin{thebibliography}{00}

\bibitem{ray} Ray, D., \textit{Development Economics}, Princeton: Princeton University Press, 1998.

\bibitem{willianson} Willianson, J.G., \textit{Handbook of developments economics}, Elsevier, Oxford, 1998.

\bibitem{harristodaro} Harris, J.R., Todaro, M.P. American Economic Review {\bf{60}} (1970) 126.

\bibitem{todaro} Todaro, M.P., American Economic Review {\bf{59}} (1969) 138.

\bibitem{bardhan} Bardhan, P., Udry, C., \textit{Development Microeconomics}. Oxford: Oxford University, 1999.

\bibitem{ranis} Ranis, G., \textit{Handbook of developments economics}, Elsevier, Oxford, 1998.

\bibitem{day} Day, R. H. et al, The Economic Journal {\bf{97}} (1987) 940.

\bibitem{summers} Summer, L.H., American Economic Review {\bf{78}} (1988) 383.

\bibitem{romer} Romer, D., \textit{Advanced Macroeconomics}, McGraw-Hill, New York, 1996.

\bibitem{brock} Brock, William A., Durlaf, Steven N. Review of Economic Studies {\bf{68}} (2001) 235.

\bibitem{durlauf} Durlauf, Steven N., \textit{The economy as an evolving complex system II}, Addison-Wesley, Santa Fe, 1997.

\bibitem{gustavo} Freitas, G.G., Master Thesis, Instituto de Pesquisas Econ\^omicas, S\~ao Paulo University, 2003.

\bibitem{thadeu} Stauffer, D., Penna, T.J.P., Physica A {\bf{256}} (1998) 284.

\bibitem{mankiwromer} Mankiw, N.G., Romer, D., \textit{New Keneysian Economics}, Vol II, MIT Press, 1991.

\end{thebibliography}
\end{document}